\documentclass[osajnl,preprint,showpacs,groupedaddress,superscriptaddress]{revtex4}

\usepackage[dvips]{graphicx}

\begin{document}
\title{Modulation induced frequency shifts in a CPT-based atomic
  clock}

\author{D.\ F.\ Phillips}
\affiliation{Harvard-Smithsonian Center for Astrophysics, Cambridge,
Massachusetts, 02138}
\author{I.\ Novikova}
\affiliation{Harvard-Smithsonian Center for Astrophysics, Cambridge,
Massachusetts, 02138}
\author{C.\ Y.-T. Wang}
\affiliation{Harvard-Smithsonian Center for Astrophysics, Cambridge,
Massachusetts, 02138}
\author{M.\ Crescimanno}
\affiliation{Department of Physics and Astronomy, Youngstown State
  University, Youngstown, Ohio 44555}
\author{R.\ L.\ Walsworth}
\affiliation{Harvard-Smithsonian Center for Astrophysics, Cambridge,
Massachusetts, 02138}
\begin{abstract}
  
  We investigate systematic errors associated with a common modulation
  technique used for phase sensitive detection of a coherent
  population trapping (CPT) resonance. In particular, we show that
  modification of the CPT resonance lineshape due to the presence of
  off-resonant fields leads to frequency shifts which may limit
  the stability of CPT-based atomic clocks. We also demonstrate that
  an alternative demodulation technique greatly reduces these effects.
\end{abstract}


\date{\today}

\maketitle


Coherent Population Trapping (CPT)~\cite{Arimondo96} has recently been
applied to the development of small stable clocks with observed
fractional frequency stability (Allan deviation) of better than
$10^{-11} \ \tau^{-1/2}$ for averaging times, $\tau$, around 100
seconds~\cite{Vanier03,Merimaa03,Kitching00,knappe01}.  CPT clocks
have potential advantages relative to traditional intensity optical
pumping clocks (which typically employ an optical and microwave double
resonance technique)~\cite{Vanier03} including the possibility of
substantial miniaturization without degradation of
performance~\cite{kitching02,knappe04}.
However, various mechanisms may degrade the frequency stability of CPT
clocks below that theoretically expected from the observed
signal-to-noise ratio. One key mechanism for such degradation is
conversion of FM laser noise to AM noise in the detected CPT
transmission signal, due to the optical absorption profile when a
laser with a large linewidth is used~\cite{Kitching01,Camparo97}. Here
we demonstrate a second important degradation mechanism: a widely-used
slow phase-modulation technique leads to shifts of the clock frequency
in the presence of asymmetries in the CPT resonance.  We also show
that a straightforward variation of this modulation technique
(\emph{i.e.}, use of third-harmonic demodulation) can eliminate much
of the systematic effect on the clock frequency.

A CPT clock employs two optical fields that are nominally resonant
with electronic transitions in alkali atoms such as Rb or Cs, with the
frequency difference between the optical fields being equal to the
hyperfine splitting of the alkali's electronic ground-state.
Initially, these two fields optically pump the atoms into a
non-interacting coherent superposition of two hyperfine states (a
``dark state'')~\cite{Arimondo96}. The long relaxation time of the
electronic ground-state leads to enhanced transmission of the optical
fields in a narrow resonance around the difference frequency of the
two optical fields. The center frequency of this resonance serves as the
CPT clock frequency; the width of the resonance is determined by the
decoherence rate of the dark state.

\begin{figure}
\begin{center}
\includegraphics[width=0.5\textwidth]{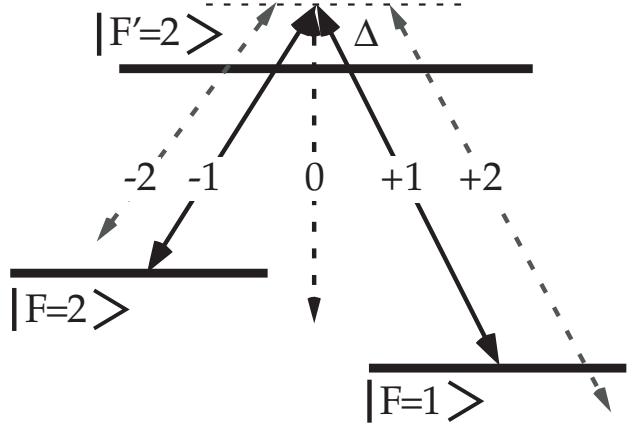}
\end{center}
\caption{
  Energy level diagram for a three-level atom coupled via two
  near-resonant fields: the +1 and -1 sidebands from a modulated
  carrier laser field, with one-photon detuning $\Delta$. Also shown
  are non-resonant fields (the carrier, $0$, and the +2 and -2
  sidebands) which can produce shifts and distortions in the CPT
  resonance. For the current experiments using ${}^{87}$Rb, the two
  lower levels correspond to the ground-electronic-state hyperfine
  levels F=2 and F=1; and the upper level corresponds to
  $5{}^2P_{1/2}$ F${}^\prime = 2$. }
\label{f.3leveloffres}
\end{figure}

In the typical manifestation of a CPT clock, a current-modulated diode
laser produces the two resonant optical fields
(Fig.~\ref{f.3leveloffres}).
Such a laser, with an optical frequency $\omega_\mathrm{opt}$ and
modulated at a microwave frequency $\omega_\mu$, generates an electric
field,
\begin{eqnarray}
\mathcal{E} & = & \mathcal{E}_0 \cos{\omega_\mathrm{opt}
    t} +  \\ \nonumber
 & &   \mathcal{E}_{-1} \cos{\left(\omega_\mathrm{opt} -
    \omega_\mu \right) t} +
   \mathcal{E}_{+1} \cos{\left(\omega_\mathrm{opt} +
    \omega_\mu \right) t}  + \\ \nonumber
& &  \mathcal{E}_{-2} \cos{\left(\omega_\mathrm{opt} -
    2 \omega_\mu \right) t}  +
   \mathcal{E}_{+2} \cos{\left(\omega_\mathrm{opt} +
    2 \omega_\mu  \right) t}  + \\ \nonumber
& & \ldots
\label{e.currentMod}
\end{eqnarray}
where $\mathcal{E}_i$ are the amplitudes of the various frequency
components of the electric field. The corresponding Rabi frequencies
for the atomic transitions driven by the fields are
$\Omega_i=d_i\mathcal{E}_i/(2\hbar)$ where $d_i$ is the atomic
transition dipole moment.
If the modulation frequency is equal to half the ground-state
hyperfine splitting ($\omega_\mu=\Delta_\mathrm{hfs}/2$) and the laser
carrier ($\omega_\mathrm{opt}$) is tuned midway between the electronic
transition frequency for the ground-state hyperfine sublevels, then
the first-order ($\pm1$) sidebands are simultaneously resonant with
the electronic transitions for the two hyperfine sublevels, and a
maximum optical transmission is observed.  However, additional
off-resonant fields are also created by the laser current-modulation.
(The carrier, $\Omega_0$, and the second-order sidebands,
$\Omega_{\pm2}$, are generally the most significant off-resonant
fields.) Even though these fields are far-detuned from the atomic
resonances, they induce an AC Stark shift, which depends on both the
optical field frequency and intensity, and causes a relative shift in
the atomic levels (a ``light shift'' in the CPT clock frequency). This
sensitivity of the CPT resonance to laser detuning and intensity can
limit the stability of a CPT clock.  Fortunately, with careful choice
of current-modulation index, AC Stark shifts from different
off-resonant fields can be arranged to cancel one
another~\cite{Vanier03}.

Phase sensitive detection is often used to provide a sensitive
feedback signal for the local oscillator (\emph{e.g.}, a quartz
crystal) which is typically locked to the CPT resonance to make a
functioning clock. Typically, a ``slow'' ($<1$ kHz) phase modulation
is superimposed on the microwave source, leading to a modification of
the phase of the microwave drive:
\begin{equation}
\omega_{\mu}t \rightarrow \omega_{\mu}t + \epsilon
\sin{\left(2\pi f_m t\right)}
\label{f.slowphase}
\end{equation}
where $f_m$ is the modulation frequency and $\epsilon$ is the
modulation index. After demodulation the (approximately) symmetric CPT
transmission resonance is transformed into an (approximately)
antisymmetric dispersion-like signal.  Near an \emph{exactly
  symmetric} CPT resonance, this dispersion-like signal is
proportional to the frequency difference between the local oscillator
and the center of the CPT resonance, and thus the local oscillator can
be locked exactly to the CPT resonance by measuring the microwave
frequency ($\omega_\mu$) corresponding to the zero-crossing of the
dispersion-like signal.  However, any asymmetry in the CPT
transmission resonance lineshape shifts the zero-crossing of the
dispersion-like signal, with a magnitude dependent upon the slow
phase-modulation parameters.  Instability in these modulation
parameters will thus induce instability in the CPT clock microwave
frequency. Here we present a detailed experimental study of such
shifts, and demonstrate that the main contribution is made by the
far-detuned optical carrier, $\Omega_0$, (designated as field ``0'' in
Fig.~\ref{f.3leveloffres}).  Note that in principle, the slow
phase-modulation and the resultant systematic effects on CPT clock
performance may be eliminated. For example, the microwave signal that
modulates the laser may be produced by direct feedback from a fast
output photodetector~\cite{strekalov04}.

\begin{figure}
\begin{center}
\includegraphics[width=\textwidth]{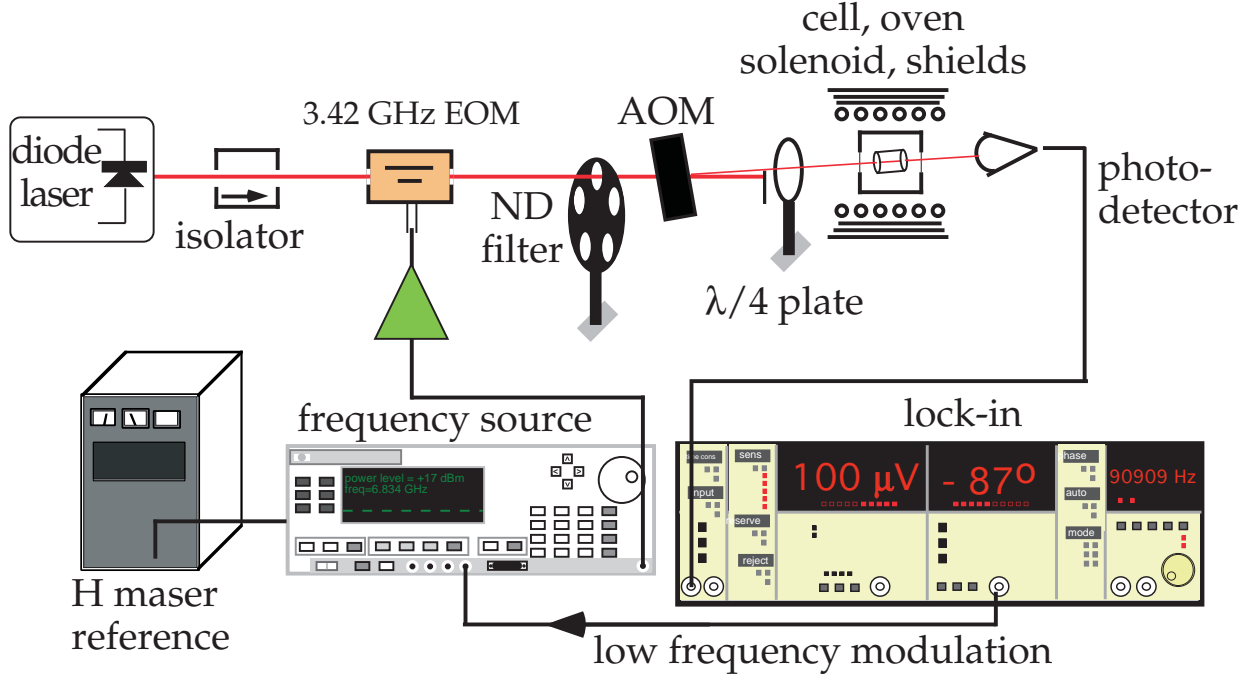}
\end{center}
\caption{Schematic of the apparatus.}
\label{f.schem}
\end{figure}
\begin{figure}
\begin{center}
\includegraphics[width=\textwidth]{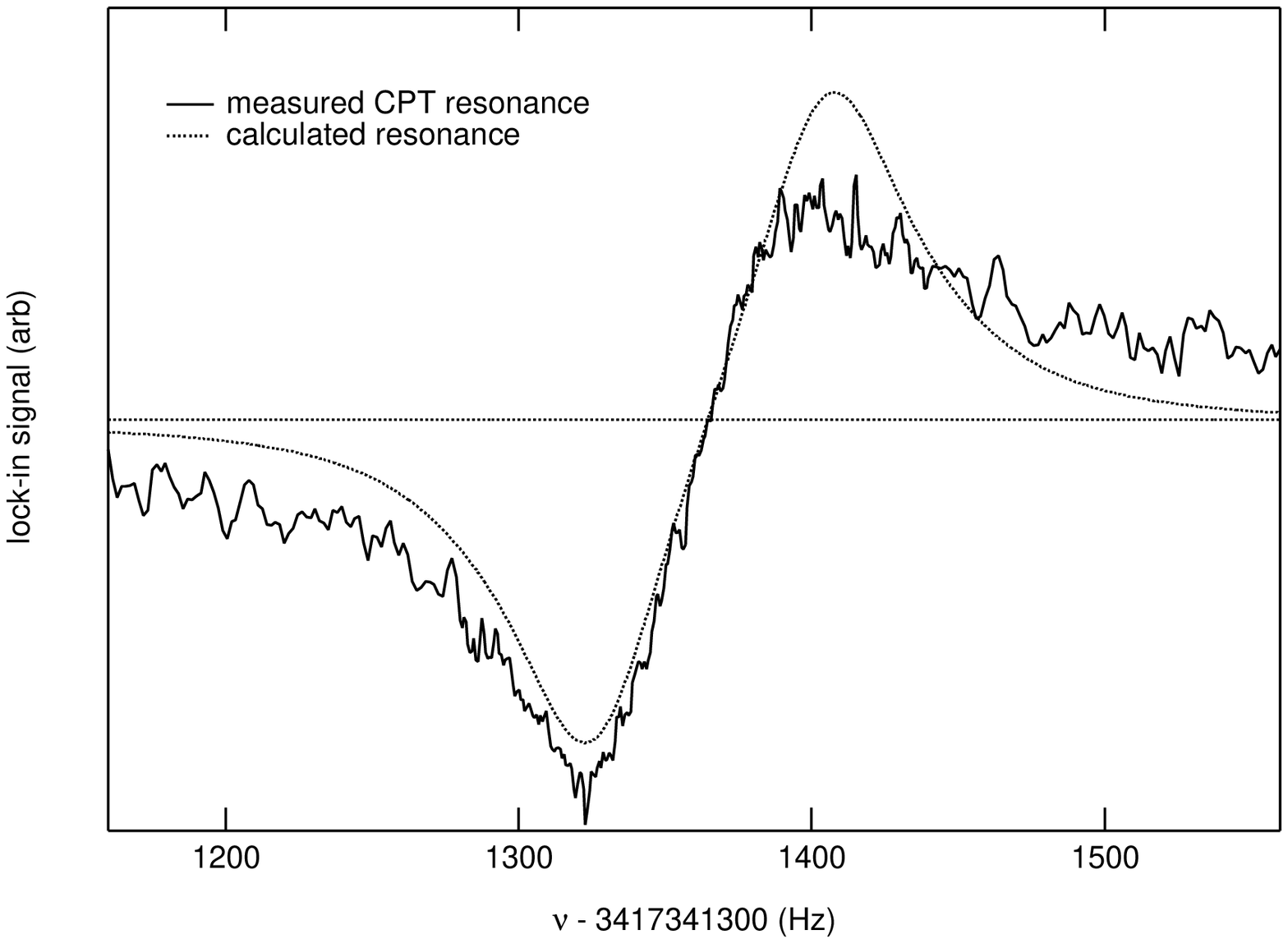}
\end{center}
\caption{Typical dispersive-like CPT resonance measured with a slow
  phase-modulation frequency $f_m=230$ Hz and index $\epsilon=0.6$
  (solid line). Also shown is a numerical calculation of the line
  shape expected from an ideal three-level system and just two
  near-resonant optical fields, for our observed Rabi frequencies and
  slow phase-modulation index (dotted line). [The width and center
  frequency of the calculated resonance were scaled to match the
  measured resonance.]  Note the asymmetry of the measured resonance
  compared to the calculation for the ideal three-level/two-field
  system.  }
\label{f.sampleres}
\end{figure}

Fig.~\ref{f.schem} shows a schematic of our experimental set-up.  We
derived the two optical fields needed for the CPT clock by phase
modulating the output from an external cavity diode
laser~\cite{newfocuslaser} tuned in the vicinity of the $D_1$ line of
Rb ($5^2S_{1/2}\longrightarrow5^2P_{1/2}$, $\lambda \simeq 795$~nm).
An electro-optic modulator (EOM)~\cite{newfocuseom} produced the phase
modulation of the optical field at half of the ground-state hyperfine
frequency of ${}^{87}$Rb ($\Delta_\mathrm{hfs}\simeq 6.8$~GHz). With
the microwave power available to the EOM in these measurements
(roughly 1 watt), approximately $15\%$ of the incident laser power was
transferred to all sidebands, with the remainder residing in the
off-resonant carrier.  Ideally, the EOM should produce equal
amplitudes in the $\pm 1$ sidebands, but due to a slight misalignment
of the input polarization, the $+1$ sideband was 1.3 dB larger than
the $-1$ sideband, producing a 35\% larger Rabi frequency
($\Omega_{+1}$) for the $|F=1\rangle\longrightarrow|F^\prime=2\rangle$
transition, than the Rabi frequency ($\Omega_{-1}$)
$|F=2\rangle\longrightarrow|F^\prime=2\rangle$ transition.  Following
the EOM, all the optical fields were attenuated by a neutral density
(ND) filter and an acousto-optic modulator (AOM) to approximately 10
$\mu$W total power, circularly polarized using a quarter wave
($\lambda/4$) plate, and then weakly focused to a diameter of about
$0.8$~mm as they passed through a Rb vapor cell.

The vapor cell was placed inside three-layers of high permeability
magnetic shielding to screen out external fields, with a solenoid also
inside the magnetic shields to control the magnetic field.  The 2.5 cm
diameter, 5 cm long vapor cell contained natural abundance Rb and 5
Torr of nitrogen buffer gas. The buffer gas slowed the Rb atomic
motion through the vapor cell.  The vapor cell was thermally
stabilized using a blown-air oven at a temperature of $45\ {}^\circ$C.
The total Rb vapor pressure at this temperature corresponds to a
${}^{87}$Rb atomic density of $2 \cdot 10^{10}$ cm$^{-3}$.  Under
these conditions the optical depth of the vapor cell was approximately
one, for a weak resonant optical field.

A coupled, three-level $\Lambda$ system was formed by the two
first-order sidebands of the laser field, the two lower ${}^{87}$Rb
states $F=1$, $m_F=0$ and $F=2$, $m_F=0$ and the excited state
$F^\prime=2$, $m_{F^\prime}=1$ (Fig.\ \ref{f.3leveloffres}). We chose
these hyperfine sublevels so as to have no first-order dependence of
hyperfine transition frequency on magnetic field.  We applied a
magnetic field of 16 mG to lift Zeeman degeneracies and remove other
ground-state sublevels from two photon resonance, thus preventing
unwanted coherences from developing.

We used phase sensitive detection to convert the approximately
symmetric CPT transmission resonance of the two-photon ``clock
transition'' into a dispersive-like resonance (Fig.\ 
\ref{f.sampleres}). We applied a slow phase-modulation to the
microwave source driving the EOM, and then demodulated the
corresponding slow variations in the photodetector current using a
lock-in amplifier.  In a working CPT clock, a feedback loop locks the
external oscillator to the zero-crossing of the antisymmetric,
modulated line. Here, rather than closing the loop in the feedback
system, we measured the frequency of the zero-crossing (by fitting a
line through the central part of the dispersive-like resonance)
relative to a frequency source phase-locked to a hydrogen maser.  We
then varied system parameters such as the laser detuning
($\omega_\mathrm{opt}$), slow phase-modulation frequency ($f_m$), or
modulation index ($\epsilon$), and measured the zero-crossing as a
function of these parameters.

\begin{figure}
\begin{center}
\includegraphics[width=0.7\textwidth]{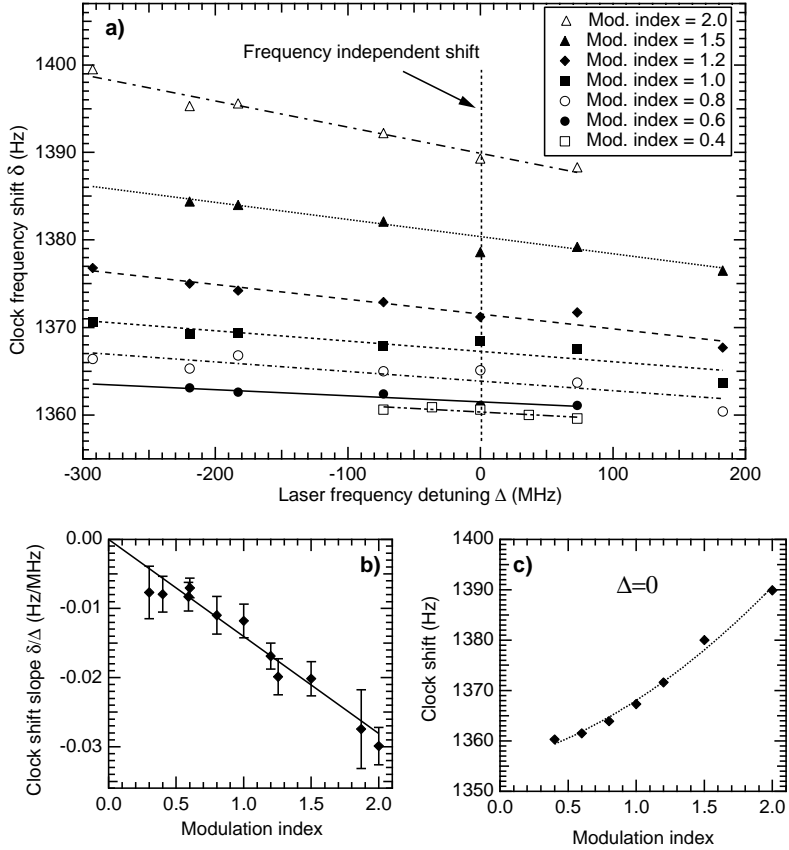}
\end{center}
\caption{a) Measured  ${}^{87}$Rb CPT clock frequency shift ($\delta$)
  as a function of detuning ($\Delta$) of the laser carrier frequency
  and resonant sidebands from $F^\prime=2$ resonance, for various slow
  phase-modulation indices ($\epsilon$).  Here, a zero frequency shift
  ($\delta=0$) corresponds to the free-space ${}^{87}$Rb hyperfine
  frequency.  The large offset of $\delta\approx1360$ Hz is due to
  the nitrogen buffer gas pressure shift.
  b) Dependence of the clock frequency on laser detuning
  ($\delta/\Delta$), determined from the slope of each individual line
  on plot (a), as a function of the slow phase-modulation index.
  c) Measured laser-frequency-independent shift, at $\Delta=0$, as a
  function of the slow phase-modulation index (see vertical line in
  plot (a)).  All data were taken at a slow phase-modulation (lock-in)
  frequency $f_m=69$ Hz.  In graphs (a) and (c),
  measurement uncertainties were comparable to the size of the symbols
  shown.}
\label{f.shift_vs_detuning}
\end{figure}

Drifts in the zero-crossing will directly lead to changes in the clock
frequency and thus degrade the frequency stability of the CPT clock.
To characterize such variations in the clock frequency ($\delta$), we
measured the dependence of the zero-crossing on the laser carrier
detuning ($\Delta$), for various slow phase (\emph{i.e.}, lock-in)
modulation parameters.  (Note that $\Delta$ was also the detuning of
the first order sidebands from the center of the Doppler broadened
absorption resonance.)  Fig.\ \ref{f.shift_vs_detuning}a shows the
measured clock frequency shift as a function of laser detuning for
various slow phase-modulation (lock-in) indices at one fixed laser
modulation index (EOM power).  We find that near the center of the
Doppler-broadened atomic transition, the clock frequency is
proportional to the laser detuning (see the linear fits in Fig.\ 
\ref{f.shift_vs_detuning}a), with a slope $\delta/\Delta$ that
increases linearly with slow phase-modulation index (at fixed slow
phase-modulation frequency) as shown on Fig.\ 
\ref{f.shift_vs_detuning}b.  We observed similar behavior upon
changing the slow phase-modulation frequency at fixed slow
phase-modulation index.  These shifts are significant in comparison to
the desired frequency stability of a CPT clock, with fractional clock
frequency sensitivity to laser detuning of order $10^{-11}$/MHz. As
shown in Fig.\ \ref{f.shift_vs_detuning}c, we also observed a
dependence of the clock frequency on slow phase-modulation index,
\emph{with the laser fields tuned to resonance}, \emph{i.e.}, with
$\Delta=0$.  This ``laser-frequency-independent'' shift is of order
$10^{-8}$ (fractionally) for 100\% changes in the modulation index.
Good CPT clock frequency stability therefore requires high stability
of the slow phase-modulation frequency and amplitude, even in the
absence of laser frequency variation.

\begin{figure}
\includegraphics[width=\textwidth]{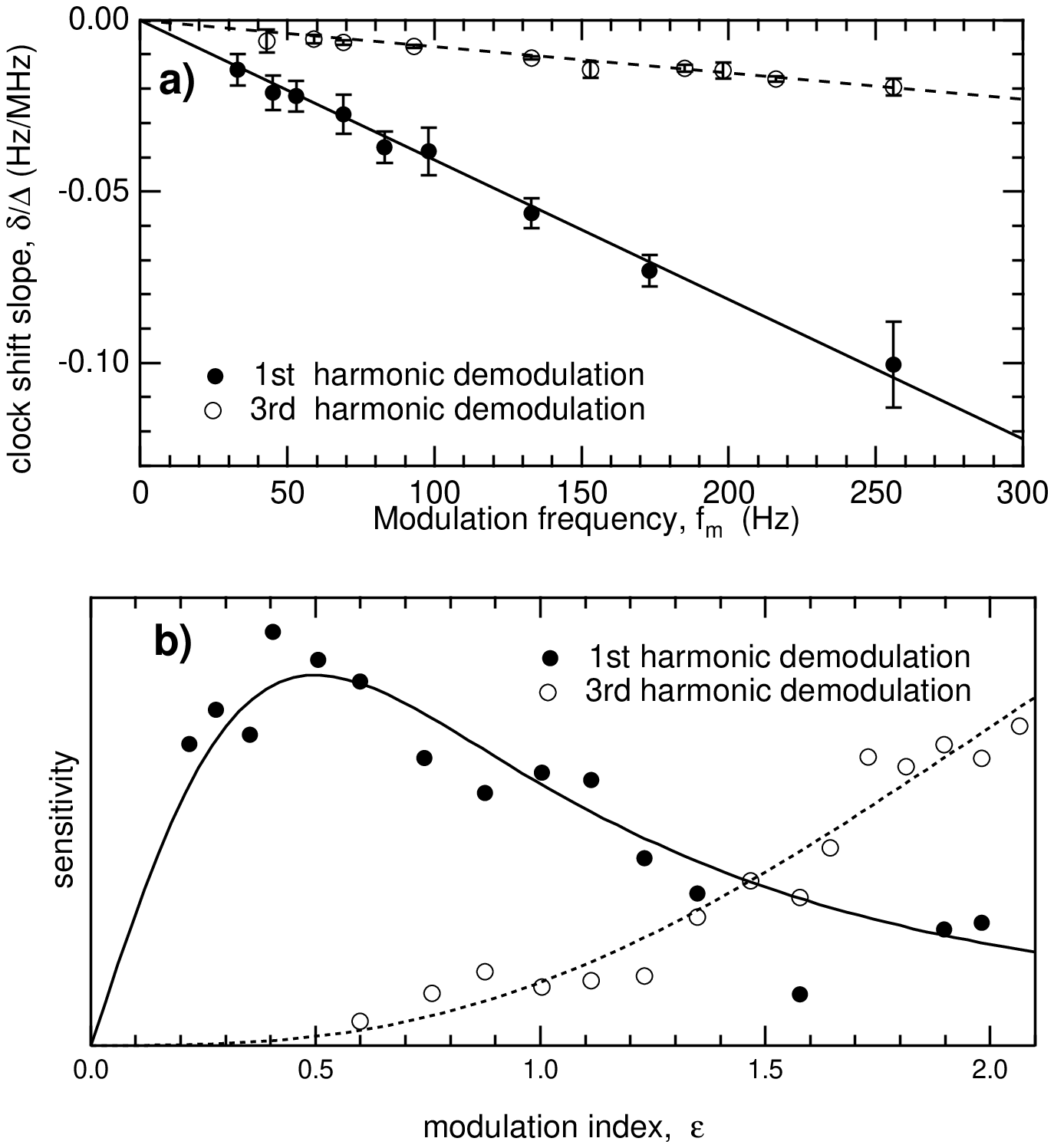}
\caption{(a) Clock frequency dependence on laser frequency
  ($\delta/\Delta$) as a function of slow phase-modulation frequency
  ($f_m$) for both first and third harmonic demodulation.  The
  modulation index $\epsilon=1.87$ for both data sets.  (b) Dependence
  of the clock transition measurement sensitivity (defined in text) to
  the slow phase-modulation index ($\epsilon$) for $f_m=153$ Hz.  A
  much larger slow phase-modulation index is required for optimal
  sensitivity in third harmonic modulation.  The lines are the results
  of a fit to the sensitivity expected for an ideal Lorentzian
  lineshape.}
\label{f.thirdHarm}
\end{figure}

Light shifts (\emph{i.e.}, AC Stark  shifts) associated with unequal
intensities of the first-order laser
sidebands~\cite{Knappe03,Taichenachev03} are expected to scale
linearly with the laser detuning:
\begin{equation} \label{dark_res_max}
\delta \propto -\Delta\frac{|\Omega_{+1}|^2-|\Omega_{-1}|^2}{\gamma^2},
\end{equation}
where $\Omega_{\pm1}$ are the Rabi frequencies of the resonant laser
sidebands and $\gamma$ is the relaxation rate of the excited state.
The data in Fig.~\ref{f.shift_vs_detuning}a exhibit this linear
scaling of $\delta$ with $\Delta$.  However, this simple light-shift
mechanism does not account for the observed dependence of the clock
frequency on the slow phase-modulation index (see Fig.\ 
\ref{f.shift_vs_detuning}b and \ref{f.shift_vs_detuning}c).  The above
light-shift expression assumes that the CPT resonance remains
symmetric near the center, only shifting due to the Rabi frequency
difference.  However, as shown in Fig.\ \ref{f.shift_vs_detuning}, the
clock frequency shift ($\delta$) depends on the modulation index
($\epsilon$) indicating that the CPT resonance is asymmetric.  [A
similar dependence of $\delta$ on the slow phase-modulation frequency
($f_m$) was also observed (Fig.\ \ref{f.thirdHarm}a).]  Additionally,
these dependencies do not vanish at $\Delta=0$, indicating that other
frequency shift mechanisms are present.

\begin{figure}
\includegraphics[width=\textwidth]{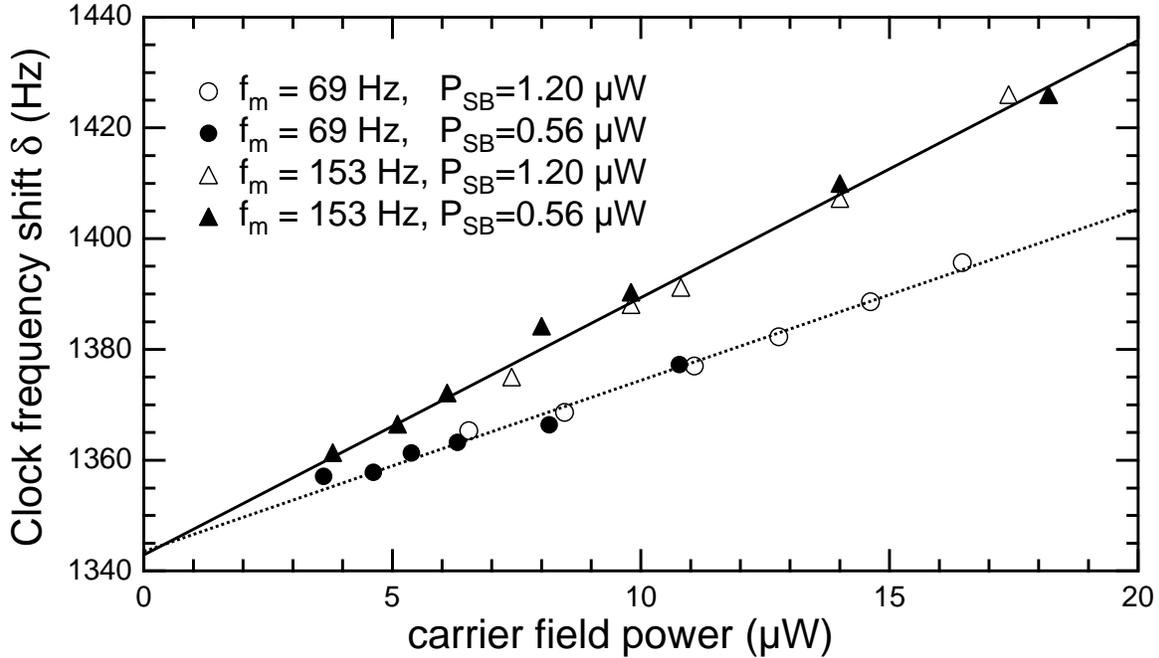}
\caption{Measured dependence of the CPT clock frequency on carrier 
  field power at one-photon resonance ($\Delta=0$), for two slow
  phase-modulation frequencies and two total powers in the first-order
  sidebands.  (Uncertainties in the measured clock frequencies are
  approximately equal to the size of the points.) Linear fits are
  shown for all data points at each of the two modulation
  frequencies.}
\label{f.carrierDep}
\end{figure}

One such mechanism is the light shift caused by the off-resonant
carrier field.  As with the Rabi frequency imbalance described above,
this light shift is usually expected to cause a simple shift in the
clock frequency but no distortion of the lineshape.  The shift is
expected to be
$\simeq 2|\Omega_{0}|^2/\Delta_0$, where $\Omega_{0}$ is the carrier
field Rabi frequency and $\Delta_0=\Delta_\mathrm{hfs}/2$ is the
magnitude of the carrier's detuning from both transitions in the
$\Lambda$-system for $\Delta=0$. Without a distortion of the
lineshape, no dependence of the clock frequency shift ($\delta$) on
the slow phase-modulation parameters should exist.
Fig.\ \ref{f.carrierDep} shows the measured clock frequency shift as a
function of laser carrier field power ($\sim|\Omega_0|^2$) at two slow
phase-modulation frequencies and two total powers in the first-order
sidebands, all for $\Delta=0$.  In addition to the usual light shift,
we observe a dependence on the slow phase-modulation frequency of
approximately 20~mHz/$\mu$W/Hz.
Consistent with Eq.\ (\ref{dark_res_max}) at $\Delta=0$, we do not
find a dependence of $\delta$ on the first-order sideband power,
$|\Omega_{\pm1}|$.
From these measurements, we conclude that the interaction of the
strong, off-resonant carrier field of the modulated laser not only
shifts the clock frequency, but also modifies the transmission
resonance lineshape.

\begin{figure}
\includegraphics[width=\textwidth]{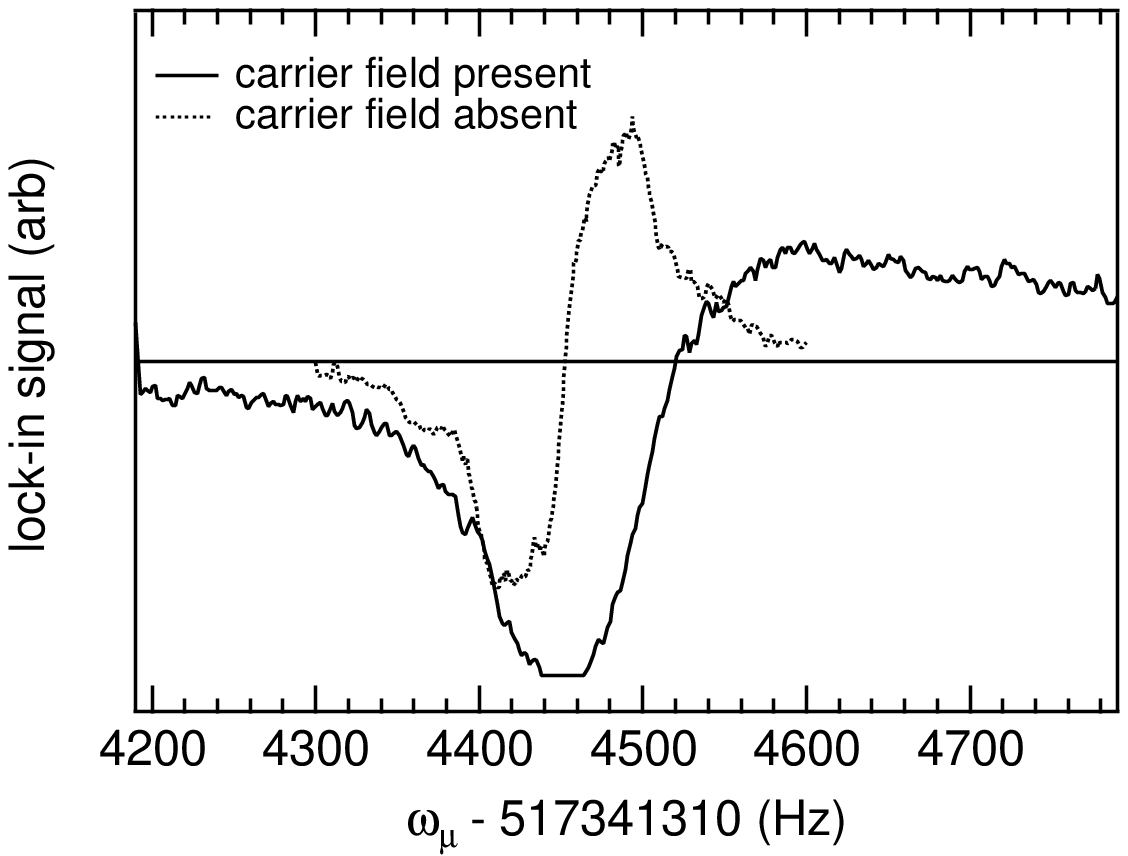}
\caption{CPT resonances with (solid line) and without (dotted line)
  the carrier field present. Both data sets were taken with $f_m=98$
  Hz and $\epsilon\approx0.6$ and with a vapor cell containing
  isotopically enriched ${}^{87}$Rb and 22 Torr of Ne buffer
  gas~\protect\cite{FPsetup}.}
\label{f.FPcleanRes}
\end{figure}

To further study the dependence of the clock frequency on the carrier
field, we inserted a Fabry-Perot cavity between the EOM and the vapor
cell in a slightly altered experimental setup~\cite{FPsetup}. This
cavity had a free spectral range of 1.37 GHz ($1/5$ of the ${}^{87}$Rb
hyperfine splitting) and was tuned such that it allowed the
transmission of the two first-order sidebands of the laser while
rejecting the carrier frequency field. A more symmetric output signal
was detected (see Fig.~\ref{f.FPcleanRes}), and the dependence of the
clock frequency shifts on the slow phase-modulation parameters
($\epsilon$ and $f_m$) were reduced by at least an order of magnitude.

We explored alternative techniques for slow phase-modulation with the
goal of reducing the dependence of the CPT clock frequency on system
parameters. For example, we investigated demodulation using the third
harmonic of the slow phase-modulation, a technique that is known to
compensate for linear asymmetry in an underlying resonance
\cite{Walls87,Marchi87}.  We found that the clock frequency for both
first- and third-harmonic demodulation has a linear dependence on the
slow modulation frequency (Fig.\ \ref{f.thirdHarm}a).  However, with
third-harmonic demodulation, the slope of this dependence is a factor
of five smaller, reducing the sensitivity of the clock frequency to
changes in either the laser carrier frequency (\emph{e.g.}, Fig.\ 
\ref{f.shift_vs_detuning}a) or the properties of the slow
phase-modulation source (Fig.\ \ref{f.shift_vs_detuning}a and b).
Similarly, we found that third-harmonic demodulation reduces the
laser-frequency-independent shift by at least an order of magnitude to
fractional shifts of less than $10^{-9}$ for 100\% changes in
modulation index.

One must also consider the change in measurement sensitivity to the
clock transition when changing the slow phase-modulation technique. We
define the measurement sensitivity as the ratio of the fitted central
slope of the dispersive-like resonance divided by the RMS fluctuations
in the fit residuals.  Optimal measurement sensitivity occurs for the
slow phase-modulation index and frequency that provide maximum signal
at the photodetector in the slow phase-modulation sidebands
demodulated by the lock-in amplifier. This optimal sensitivity is
achieved when the width of the ``comb'' of slow phase-modulation
sidebands transmitted through the atomic medium is comparable to the
CPT resonance width.  At the optimal slow phase-modulation index for
first harmonic demodulation, we found that the measurement sensitivity
when using third harmonic demodulation is substantially reduced;
however, we recovered the original sensitivity by increasing the slow
phase-modulation index (Fig.\ \ref{f.thirdHarm}b).

The measurement sensitivity's dependence on modulation index is
consistent with a simple Lorentzian model.  We calculated the
dispersive curves obtained by applying slow modulation at frequency
$f_m$ and index $\epsilon$ to a Lorentzian lineshape and determined
the slope of the curve near the zero-crossing.  The lines in Fig.\ 
\ref{f.thirdHarm}b show this slope as a function of modulation index
with only the ratio of the Lorentzian resonance linewidth to the
modulation frequency and the overall amplitude adjusted to fit the
data. Note, however, that this model does not include a realistic CPT
lineshape or the effects of off-resonant fields and thus does not
reproduce the data in Fig.\ \ref{f.thirdHarm}a.

In conclusion, we have quantitatively studied systematic effects on
CPT clock frequency due to asymmetries in the two-photon CPT resonance
induced by strong off-resonant laser fields.  These asymmetries can
produce significant shifts in the CPT clock frequency when slow
phase-modulation is used to determine the center of the CPT resonance.
While lowering the frequency of the slow phase-modulation decreases
the sensitivity of the clock to such systematic frequency shifts,
technical noise reduces the signal-to-noise ratio (and hence the
clock's short-term frequency stability) at very slow phase-modulation
frequencies.

To achieve good fractional frequency stability $\sim10^{-11}$ in a CPT
clock, these systematic frequency shifts impose demanding requirements
on the stability of the elements that control the slow
phase-modulation ($f_m$ and $\epsilon$) and the laser carrier
frequency ($\omega_\mathrm{opt}$). Fortunately, there are promising
alternative modulation techniques that can mitigate the effects of CPT
lineshape asymmetry in the presence of non-resonant laser fields.  A
careful choice of the fast (microwave) modulation index allows the AC
Stark shifts from different off-resonant laser fields to cancel each
other \cite{Vanier03}. As demonstrated here, third-harmonic
demodulation of the slow phase-modulation greatly reduces the
sensitivity to asymmetry in the CPT resonance.

We thank R.\ F.\ C.\ Vessot and E.\ M.\ Mattison for technical
assistance. B.\ Murphy and J.\ Hager contributed to early aspects of
this work.  This work was supported by the Office of Naval Research
and in part by the National Science Foundation through a grant to the
Institute of Theoretical Atomic and Molecular Physics at Harvard
University and the Smithsonian Astrophysical Observatory.


\begin{thebibliography}{}
\bibitem{FPsetup} For this series of measurements, we used a vapor
  cell with isotopically enriched ${}^{87}$Rb and 22 Torr of Ne buffer
  gas. The change in buffer gas leads to both a different overall
  clock frequency shift and a narrower CPT resonance than the 5 Torr
  N${}_2$ cell we used for the other measurements we report here.
\end{thebibliography}
\end{document}